\documentclass[a4paper,UKenglish,cleveref, autoref, thm-restate]{lipics-v2021}

\usepackage{pgfplots}
\pgfplotsset{compat=1.15}

\title{ARIA - An Agentic Framework for Autonomous Testing of Infotainment Systems} 

\author{António Azevedo}{Critical Techworks, Portugal \and Faculty of Engineering, University of Porto, Portugal} {up202108689@up.pt}{}{}

\author{Bruno Lima}{LIACC, Faculty of Engineering, University of Porto, Portugal \and \url{http://www.brunolima.info}} {brunolima@fe.up.pt}{https://orcid.org/0000-0003-2572-047X}{}

\author{João Pascoal Faria}{INESC TEC, Faculty of Engineering, University of Porto, Portugal}{jpf@fe.up.pt}{https://orcid.org/0000-0003-3825-3954}{}

\authorrunning{A. Azevedo, B. Lima and J. P. Faria} 

\Copyright{António Azevedo, Bruno Lima and João P. Faria} 


\category{Software Engineering in Practice Track Paper} 

\relatedversion{} 

\nolinenumbers 

\EventEditors{Robert Feldt, Maria Paasivaara, Daniel Mendez, Stefan Wagner, and Marvin Mu\~{n}oz Bar\'{o}n}
\EventNoEds{5}
\EventLongTitle{20th International Symposium on Empirical Software Engineering and Measurement (ESEM 2026)}
\EventShortTitle{ESEM 2026}
\EventAcronym{ESEM}
\EventYear{2026}
\EventDate{October 8--9, 2026}
\EventLocation{Munich, Germany}
\EventLogo{}
\SeriesVolume{394}
\ArticleNo{82}

\begin{document}

\maketitle

\begin{abstract}
Modern automotive infotainment systems sit at the center of the vehicle's digital ecosystem, yet their validation still relies heavily on manual testing, a process that is time-consuming, expensive, and increasingly incompatible with the pace of agile release cycles and over-the-air software updates. Traditional scripted automation offers only a partial remedy, as it creates tight coupling between test logic and implementation details, producing brittle suites with high maintenance overhead. Existing LLM-driven testing frameworks predominantly target web or mobile applications, while employing single-agent or dual-agent architectures that overload one or two models with perception, planning, action selection, and validation simultaneously, making them prone to hallucination-style failures and unproductive exploration loops when faced with the complexity of automotive infotainment interfaces. In this paper, we present ARIA (Autonomous Real-time Infotainment Assessment), a multi-agent framework that leverages Large Language Models to autonomously execute end-to-end test scenarios on Android-based infotainment systems through visual interface interaction, orchestrating a closed-loop pipeline of four specialized agents per interaction step, complemented by a dedicated report-generation stage. From single-sentence natural-language scenario descriptions alone, each specifying a navigation path, an action to perform, and an expected outcome to verify, it autonomously executes the corresponding interactions on the infotainment system and produces structured reports, reproducible action scripts, and visual evidence for each step. ARIA was evaluated in an industrial setting on a physical test environment running the Android-based infotainment system of a car manufacturer, executing 30 scenarios spanning diverse system functionalities. Of the 30 scenarios, 28 (93.3\%) completed the full multi-agent pipeline and produced a verdict, while 2 terminated prematurely with execution errors. Of the 28 completed scenarios, 20 (71.4\%) matched the ground truth. ARIA detected all 5 known functional defects in the test setup, so no genuine fault was ever passed as working; the 8 false positives among completed scenarios are attributable to LLM navigation and image-interpretation limitations and to unsupported interaction gestures. These findings demonstrate that multi-agent LLM architectures can autonomously execute end-to-end infotainment test scenarios in an industrial setting, while also exposing the precision challenges that a low false-positive tolerance imposes. A single-agent baseline ablation on the same scenarios confirms the value of the multi-agent decomposition: on the first pass, before revisitation with a stronger model masks the difference, the single agent produces a substantially higher false positive rate (72.0\% versus 52.6\%) due to the conflation of navigational difficulty with system failure. We separately report first-pass and post-revisitation results and quantify token consumption, LLM-call counts, and monetary cost per scenario for both pipelines, and repeated execution of a representative subset of scenarios confirms that outcome stability correlates with scenario complexity, with fault detection remaining perfectly consistent across runs, indicating a path toward integrating visual test execution into continuous integration pipelines.
\end{abstract}

\section{Introduction}

Modern automotive infotainment systems have evolved into highly complex software platforms integrating navigation, media, connectivity, vehicle settings, and driver-assistance interfaces into a single cohesive experience \cite{Gao2024IntelligentCockpits, garzon:ivi}. As the primary interface between the driver and the vehicle's digital capabilities, the infotainment system sits at the center of an ecosystem of individually developed electronic control units \cite{jeong:infotainment, agbaje:iov}, so its failures carry disproportionate consequences, directly undermining the user experience and customer trust, particularly for premium manufacturers \cite{jeong:infotainment, yin2018automated}. Because these systems are assembled from the contributions of many teams responsible for different subsystems and UI components, ensuring they function correctly together makes end-to-end testing essential \cite{tierno_open_2017}.

This integration validation has traditionally relied heavily on manual testing: human testers follow scripted procedures on physical or virtual hardware, interacting with the system as an end user would and verifying each step \cite{tierno_open_2017, nikhade_advanced_2023}. While thorough, this is time-consuming and expensive, and defects that could have been caught early are instead discovered late, when remediation is costlier \cite{najihi:testing}. The problem is amplified by over-the-air updates and agile release cycles, where each iteration demands regression testing that scales poorly with manual execution, motivating the ``shift-left'' principle of moving quality assurance earlier in the development life cycle \cite{najihi:testing, pham_review_2022}. This burden demands a departure from traditional approaches toward intelligent, adaptive testing frameworks capable of reasoning about the system as an end user would \cite{faraji_ai-driven_2025, jha_artificial_2021}.

This work explores the use of Large Language Models (LLMs) to automate the execution of these manual test scripts. Rather than encoding test logic against internal APIs or system architecture, which creates tight coupling and fragile tests, the proposed system interacts with the infotainment UI exclusively through its visual interface, like a human user would \cite{goetz_enhancing_2025}. By interpreting screenshots and navigating from high-level task descriptions, the LLM-powered agents remain decoupled from the underlying software architecture, making the approach resilient to internal refactoring and applicable across system versions. The system employs a multi-agent architecture in which specialized agents collaborate on each scenario, handling action planning, UI element identification, interaction validation, and outcome evaluation, mirroring the structured reasoning a human tester would apply \cite{ran_guardian_2024, garlapati_ai-powered_2024}. As the testing community increasingly turns to AI-driven automation to keep pace with continuous integration, this industrial context motivates frameworks that execute end-to-end scenarios autonomously, with the long-term vision of integrating visual test execution into regression pipelines \cite{faraji_ai-driven_2025, pham_review_2022, ranapana_role_2025}.

To address this growing need, we propose ARIA (Autonomous Real-time Infotainment Assessment), a novel multi-agent framework for the automated execution and validation of end-to-end test scenarios on Android-based automotive infotainment systems. The main contributions of this work are as follows:
\begin{bracketenumerate}
\item A multi-agent architecture comprising four specialized LLM-powered agents that separates perceptual reasoning from implementation-level interaction, enabling autonomous navigation and evaluation of infotainment UIs through visual interface interaction alone. 

\item A natural-language-driven execution pipeline that accepts single-sentence scenario descriptions as input, enabling organizations to ingest and execute existing manual test scripts without modification or formalization.

\item An industrial evaluation on a physical infotainment test environment, with 28 of 30 scenarios completing the pipeline and all 5 known functional defects detected, accompanied by a systematic root cause analysis that attributes the non-correct outcomes to LLM navigation and image-interpretation limitations and to unsupported gestures in the device interface.

\item A single-agent baseline ablation demonstrating that the multi-agent decomposition reduces the first-pass false positive rate from 72.0\% to 52.6\% on the same scenario set, isolating the contribution of the architecture from that of the underlying model.

\item A detailed accounting of token consumption, LLM-call counts, and monetary cost per scenario for both pipelines, with a clear separation of first-pass from post-revisitation results, together with a repeated-execution experiment that quantifies run-to-run variance and confirms that outcome stability is high for scenarios of moderate complexity, with fault detection remaining perfectly consistent across runs.


\end{bracketenumerate}

\section{ARIA}\label{sec:aria}

\subsection{System Architecture}
\begin{figure}
    \centering
    \includegraphics[width=0.68\linewidth]{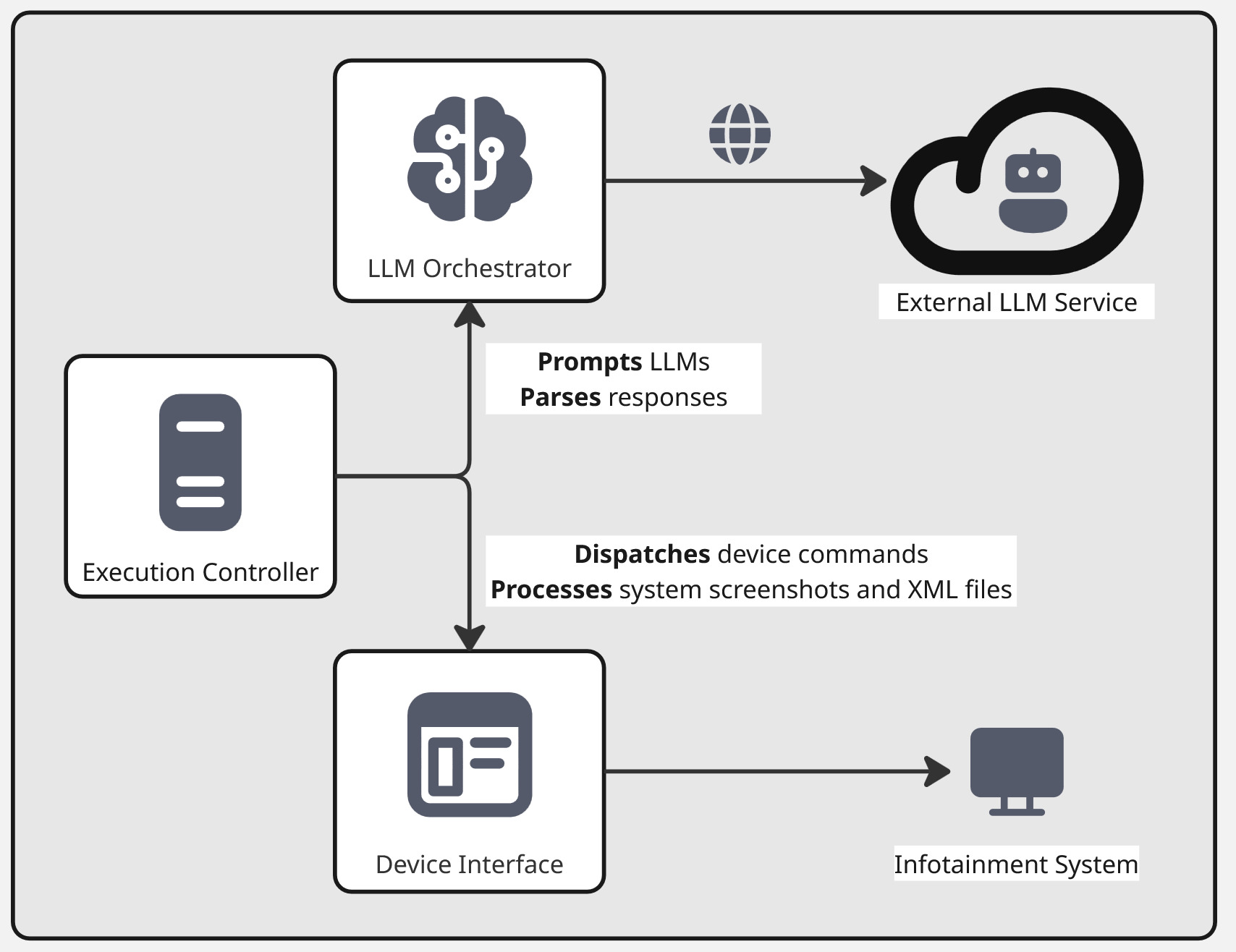}
    \caption{ARIA's architecture layers diagram}
    \label{fig:architecture}
\end{figure}

ARIA is structured as a layered system with three principal components. The \textbf{LLM orchestration layer} sits at the top, providing a standardized interface through which any supported model provider can be integrated without modifying the logic below it, and managing agent sessions and model interactions. The \textbf{execution layer}, between the LLM services and the device, implements the closed-loop multi-agent pipeline: it coordinates the four specialized agents, manages scenario state, enforces step limits and retry policies, and persists execution artifacts. The \textbf{device interface layer}, at the bottom, encapsulates all communication with the system under test through Appium, translating high-level interaction commands into coordinate-based device primitives and capturing UI state as screenshots and XML element hierarchies. This separation, reflecting architectural principles identified in the literature, ensures that the LLM provider, the agent orchestration logic, and the device communication protocol can each be changed independently, and that the framework can be adapted to a different infotainment platform by modifying only the device interface layer (Figure~\ref{fig:architecture}).

\subsubsection{LLM Service Abstraction}

Each agent is instantiated as an independent session with its own system prompt, model assignment, and message history, so that agents cannot access or be influenced by one another's context, enforcing the information boundaries discussed in Section~\ref{sec:rel_work}. The orchestration layer supports heterogeneous, plug-and-play model selection across agents. Communication follows a structured request--response pattern: prompts are multimodal messages combining text, files, and images, and responses are constrained to predefined JSON schemas through prompt engineering and post-response verification. When a response fails to conform, the layer automatically re-prompts the agent, tolerating up to five consecutive formatting failures before escalating an error, ensuring downstream agents always receive well-formed input.

\subsubsection{Multi-Agent Execution Pipeline}

\begin{figure}
    \centering
    \includegraphics[width=0.82\linewidth]{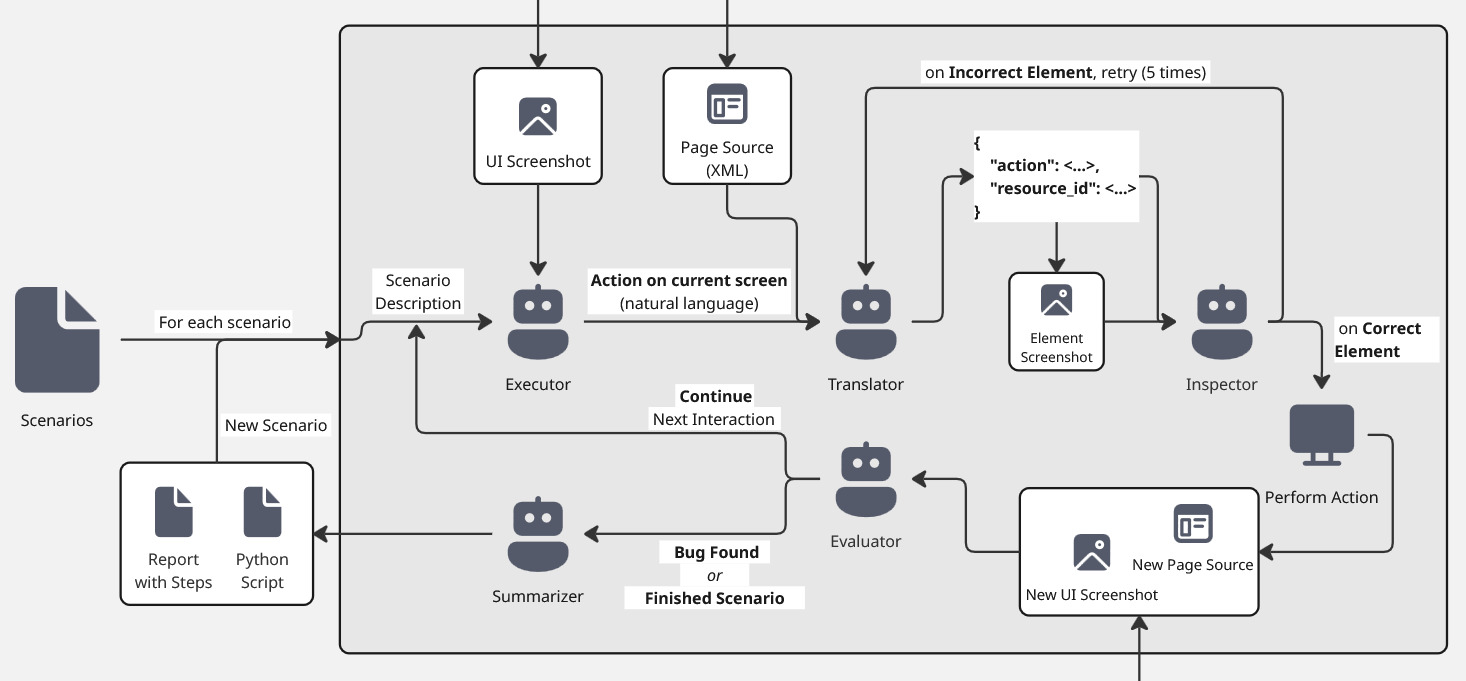}
    \caption{ARIA's standard execution diagram}
    \label{fig:aria_execution}
\end{figure}

The core of ARIA's operation is a closed-loop pipeline that iterates through four specialized agents per interaction step (Figure~\ref{fig:aria_execution}). Given a scenario described in natural language, the pipeline proceeds as follows.

The \textbf{Executor} agent receives the current screenshot and the scenario description; on its first invocation it establishes the high-level task, and on subsequent invocations it determines the next action from the updated screenshot. It is instructed to simulate human end-user behavior and operates without access to the XML element hierarchy, internal API specifications, or system documentation, basing its decisions solely on what is visually apparent on screen.

The \textbf{Translator} agent receives the high-level action together with the filtered XML page source, identifies the specific UI element that corresponds to the intended action, and produces a structured JSON object with the target's resource identifier, action type, and parameters; the system then resolves the element's bounding coordinates and interaction point. If the resource identifier does not exist in the current page source, the agent is re-prompted with an error message, up to a maximum of five retries before the scenario is terminated with an error status.

The \textbf{Inspector} agent receives the intended action and a cropped image of the screen region surrounding the element selected by the Translator, and returns a binary judgment (pass or fail) on whether the element appears semantically appropriate for the action. If the element is rejected, the Translator is re-prompted to select an alternative, providing a safeguard against incorrect element mappings before any interaction reaches the device.

Once the Inspector approves the element, the interaction is executed on the device and, after a fixed settling delay, a new screenshot is presented to the \textbf{Evaluator} agent, which compares the previous state, the action taken, and the resulting state and classifies behavior into one of three categories: \textit{Correct} (the expected change occurred and the pipeline loops back to the Executor), \textit{Finished} (the objective has been achieved and the scenario concludes successfully), or \textit{Incorrect} (an unexpected behavior indicates a potential fault and the scenario terminates). A configurable maximum step limit terminates the scenario with an error status if the objective is not reached.

An additional agent, the \textbf{Summarizer}, operates outside the main loop: after each scenario completes, regardless of outcome, it turns the full execution log into a structured JSON report containing the scenario description, the overall result (\texttt{OK}, \texttt{BUG}, or \texttt{ERROR}), an optional failure description, and a chronological list of steps, persisted to disk and reflected in the run manifest, providing both human-readable documentation and machine-parseable records.

\subsubsection{Device Interface}
Communication with the system under test is mediated through Appium, an open-source mobile automation framework \cite{appiumdocs}: ARIA connects to the infotainment unit as a remote Android device over a WebDriver-compatible API, with device discovery performed automatically via ADB \cite{androidadb}. The interface provides tap, long-tap, swipe, and text-input primitives, each parameterized by screen coordinates; text input bypasses the on-screen keyboard by injecting characters directly through ADB shell commands. A fixed three-second delay follows each interaction to let the system settle, and every executed interaction is recorded with its type, coordinates, and parameters, enabling automatic generation of standalone replay scripts.

At each decision point the interface captures two complementary representations of UI state: a screenshot, and the XML element hierarchy from Appium's page source API. The raw XML is parsed and filtered to the actionable nodes that possess both a resource identifier and bounding coordinates, retaining a minimal attribute set (\texttt{bounds}, \texttt{text}, \texttt{content-desc}, \texttt{resource-id}, \texttt{clickable}, \texttt{checkable}) and further distilled into a compact \texttt{resource-id}/\texttt{bounds} CSV for coordinate lookup. Together with the screenshot, these constitute the device-state representation distributed to the agents at each step.

\subsection{Execution Modes}

ARIA supports four execution modes, each addressing a different concern of the validation process.

\subsubsection{Standard Execution}

Standard execution processes scenarios sequentially through the full pipeline. Each scenario begins from the home screen and runs the closed loop until it completes, a fault is detected, or an unrecoverable error occurs; a run manifest with a checkpoint index records per-scenario status and enables resumption after interruptions. The system returns to the home screen between scenarios, and any scenario that ends in \texttt{BUG} or \texttt{ERROR} is automatically queued for re-visitation.

\subsubsection{Short-Circuit Replay}

\begin{figure}
    \centering
    \includegraphics[width=0.72\linewidth]{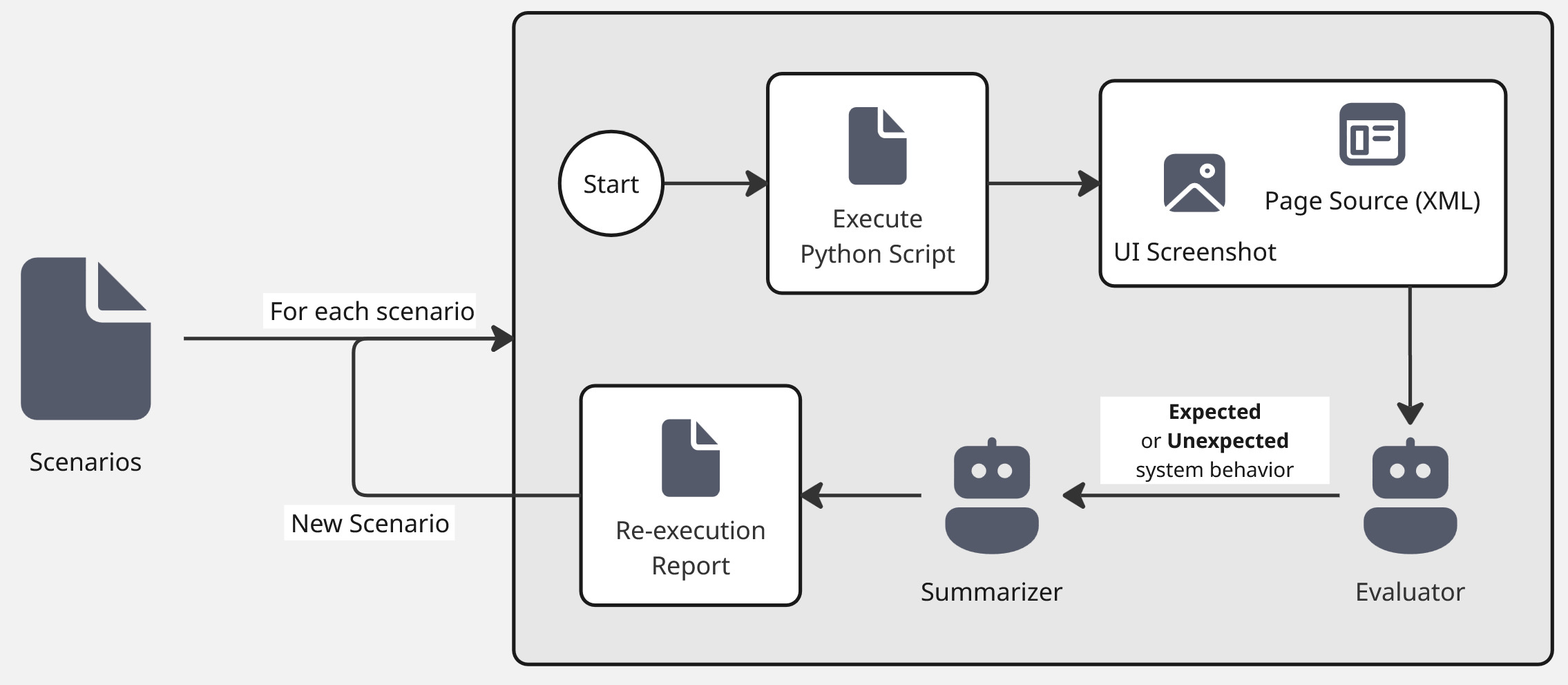}
    \caption{Short-Circuit Replay mode's execution diagram}
    \label{fig:reexecution}
\end{figure}

Short-circuit replay enables deterministic re-evaluation without the cost of the full pipeline: previously generated action scripts are replayed on the device as sub-processes, bypassing the Executor, Translator, and Inspector, and a standalone Evaluator classifies the final state against the scenario description. If the same interactions on an updated software version yield a different final state, the discrepancy is flagged, providing an efficient mechanism for detecting regressions introduced between software iterations (Figure~\ref{fig:reexecution}).

\subsubsection{Re-visitation with Stronger Models}

Re-visitation targets scenarios that previously concluded with \texttt{BUG}, \texttt{ERROR}, or \texttt{PENDING} status (which happens when the program crashes during the execution of a scenario), re-executing them through the full pipeline with a more capable model. Motivated by the observation that execution failures are often model limitations (incorrect element selections or misinterpreted screenshots) rather than genuine defects, it provides a second evaluation that reduces false positives and improves confidence in results that persist across both runs.

\subsubsection{Scenario Transformation}

Scenario transformation converts existing manual test cases into natural-language scenario descriptions without a device connection: a generation agent reformulates batches of manual test definitions into the navigation--action--verification format expected by the Executor, letting organizations feed existing test repositories into the framework without manual re-authoring.

\subsection{Simulating Human User Behavior}

A central design principle is that the Executor reasons about the system as a human user would, achieved through deliberate information restriction: it receives only the current screenshot and the natural-language scenario, with no access to the XML hierarchy, internal APIs, or documentation, forcing decisions based on visual cues, spatial layout, and interface conventions. The separation between the Executor (perceptual level) and the Translator (implementation level, resolving intentions into coordinates via the XML) prevents implementation knowledge from contaminating the user-centric perspective. As a result, the agent can encounter the same usability issues, navigational confusion, or misleading interface elements that a real user would, increasing the validity of the automated execution.

\subsection{Limitations}

Despite its capabilities, ARIA's current implementation is subject to several limitations that must be considered when interpreting results.

\textbf{Model Limitations.} Agents may reference elements that do not exist on the current screen, misidentify a visible element's function, or propose unavailable actions. The Inspector and the Translator's identifier verification mitigate this, but Executor-level mistakes (such as believing a menu is open when it is not) propagate through the pipeline before they can be caught, sometimes causing correct system behavior to be classified as an error and necessitating the re-visitation mechanism.

\textbf{Execution Speed.} Each step requires sequential invocations of four agents, each a round trip to an external LLM service, compounded by the fixed post-interaction settling delays. Per-scenario times are therefore substantially longer than scripted automation, acceptable for nightly regression runs but precluding rapid in-loop feedback during active development.

\textbf{Token Consumption.} Transmitting screenshots at every decision point yields high per-scenario token consumption, translating directly into operational cost that may constrain the scale and frequency of test campaigns.

\textbf{Interaction Pattern Constraints.} The interface supports tap, long-tap, swipe, and text input; compound gestures such as long-press-and-drag or coordinated swipes between element positions, and interactions that depend on detecting rapid transient UI changes, are not currently supported, so scenarios requiring them must be excluded or simplified.

\section{Evaluation}\label{sec:validation}

\subsection{Experimental Design}

The evaluation of ARIA was designed as an industrial case study conducted on a physical infotainment test environment running a baseline software version with known characteristics. Following the structure of comparable industrial evaluations in LLM-based test automation, the study is organized around four research questions, each targeting a distinct aspect of the framework's design, capability, and practical applicability.

\textbf{RQ1.} How can LLM-powered multi-agent architectures be designed to autonomously execute end-to-end test scenarios on automotive infotainment systems through visual interface interaction?

\textbf{RQ2.} To what extent can a vision-based, natural-language-driven testing framework reliably detect genuine system faults while minimizing false positives in an industrial infotainment validation environment?

\textbf{RQ3.} What are the primary failure modes, limitations, and architectural trade-offs involved in delegating GUI test execution to LLM agents operating through coordinate-based interaction primitives?

\textbf{RQ4.} What are the practical challenges and prerequisites for integrating LLM-based test execution frameworks into existing automotive validation workflows and continuous integration pipelines?

\subsection{Experimental Setup}

The test setup operates as a standalone unit disconnected from a vehicle, so functionalities dependent on vehicle state, sensor input, or active network services were unavailable. Core functionalities were operational except for the climate control subsystem, which was purposefully disabled to serve as a benchmark for expected faulty behavior. ARIA connected to the unit remotely via Appium, with device discovery handled through ADB.

Thirty test scenarios were selected, spanning home screen navigation, widget interaction, application launching, search operations, settings manipulation, climate controls, connectivity toggles, status bar interactions, and deep menu traversal. Scenarios requiring external dependencies (active phone connections, internet access, radio reception, navigation services, or specific account configurations) or with ambiguous verification criteria were excluded.

Execution used the standard mode with \texttt{claude-sonnet-4.6} as the default model \cite{anthropic2026claude_sonnet}, a maximum of ten steps per scenario, and five translation retries. Scenarios terminating with BUG or ERROR status were automatically revisited using \texttt{claude-opus-4.6} \cite{anthropic2026claude_opus}, with the system returning to the home screen between scenarios to ensure a consistent starting state.

\subsubsection{Model Selection}\label{sec:model_selection}

To justify the choice of LLM models, a preliminary validation study was conducted using a representative subset of six scenarios selected from the full evaluation set, spanning scenarios of varying complexity including simple status bar checks, connectivity toggles, settings navigation, application search, and a known system defect. Five candidate models were evaluated: \texttt{claude-sonnet-4.6}, \texttt{claude-opus-4.6}, \texttt{gpt-4.1}, \texttt{gpt-5.3-codex}, and \texttt{gpt-5.4}. All models were evaluated under identical conditions using the same multi-agent pipeline configuration. The ground truth for these scenarios was established in advance: five scenarios expected to classify as OK and one expected to classify as BUG. Table~\ref{tab:model-selection} summarizes the results.

\begin{table}[htbp]
\centering
\begin{tabular}{lccccc}
\textbf{Model} & \textbf{Completed} & \textbf{Correct} & \textbf{FP} & \textbf{FN} & \textbf{Cost (est.)} \\
\hline
\texttt{claude-sonnet-4.6} & 4/6 & 4/4 & 0 & 0 & \$8.28 \\
\texttt{claude-opus-4.6} & 5/6 & 4/5 & 1 & 0 & \$28.56 \\
\texttt{gpt-4.1} & 1/6 & 1/1 & 0 & 0 & \$0.00* \\
\texttt{gpt-5.3-codex} & 0/6 & --- & --- & --- & \$4.24 \\
\texttt{gpt-5.4} & 2/6 & 1/2 & 1 & 0 & \$4.08 \\
\end{tabular}
\caption{Model selection validation results across 6 representative scenarios. FP = false positives among completed scenarios; FN = false negatives among completed scenarios (the known defect completed but misclassified as OK). Cost estimated from premium request counts. * GPT-4.1 uses a separate billing model and is considered free.}
\label{tab:model-selection}
\end{table}

The OpenAI model family exhibited substantially lower pipeline completion. \texttt{gpt-5.3-codex} completed none of the six scenarios, terminating with errors due to persistent failures in producing JSON conforming to the required agent schemas, while \texttt{gpt-4.1} and \texttt{gpt-5.4} completed only one-sixth and one-third respectively, in both cases error-terminating on the buggy scenario itself, so neither ever produced a verdict on the known defect, struggling with the structured multi-turn interaction the Translator requires.

Among the Anthropic models, \texttt{claude-sonnet-4.6} achieved perfect classification on completed scenarios (4/4, zero false positives) at the lowest cost (\$8.28 across 207 LLM calls), whereas \texttt{claude-opus-4.6} completed one additional scenario (5/6) but introduced a false positive at roughly 3.4$\times$ the cost (238 calls, \$28.56); both correctly detected the known defect. These results motivated selecting \texttt{claude-sonnet-4.6} as the primary execution model for its accuracy, cost, and reliability trade-off, and \texttt{claude-opus-4.6} as the revisitation model for its higher completion rate, acceptable since revisitation applies only to the subset needing re-execution. Open-weight models were not evaluated, as their visual reasoning at the time fell short of the multimodal demands of reliable element identification and behavioral classification.

\subsection{Results}

To establish ground truth, manual inspection was performed on each of the 30 evaluated scenarios after execution, determining the expected outcome from the known state of the infotainment software and the capabilities available in the test environment, classifying each result as a correct assessment or a misclassification, and documenting the root cause of any discrepancy.

\begin{figure}[htbp]
    \centering
    \includegraphics[width=1\textwidth]{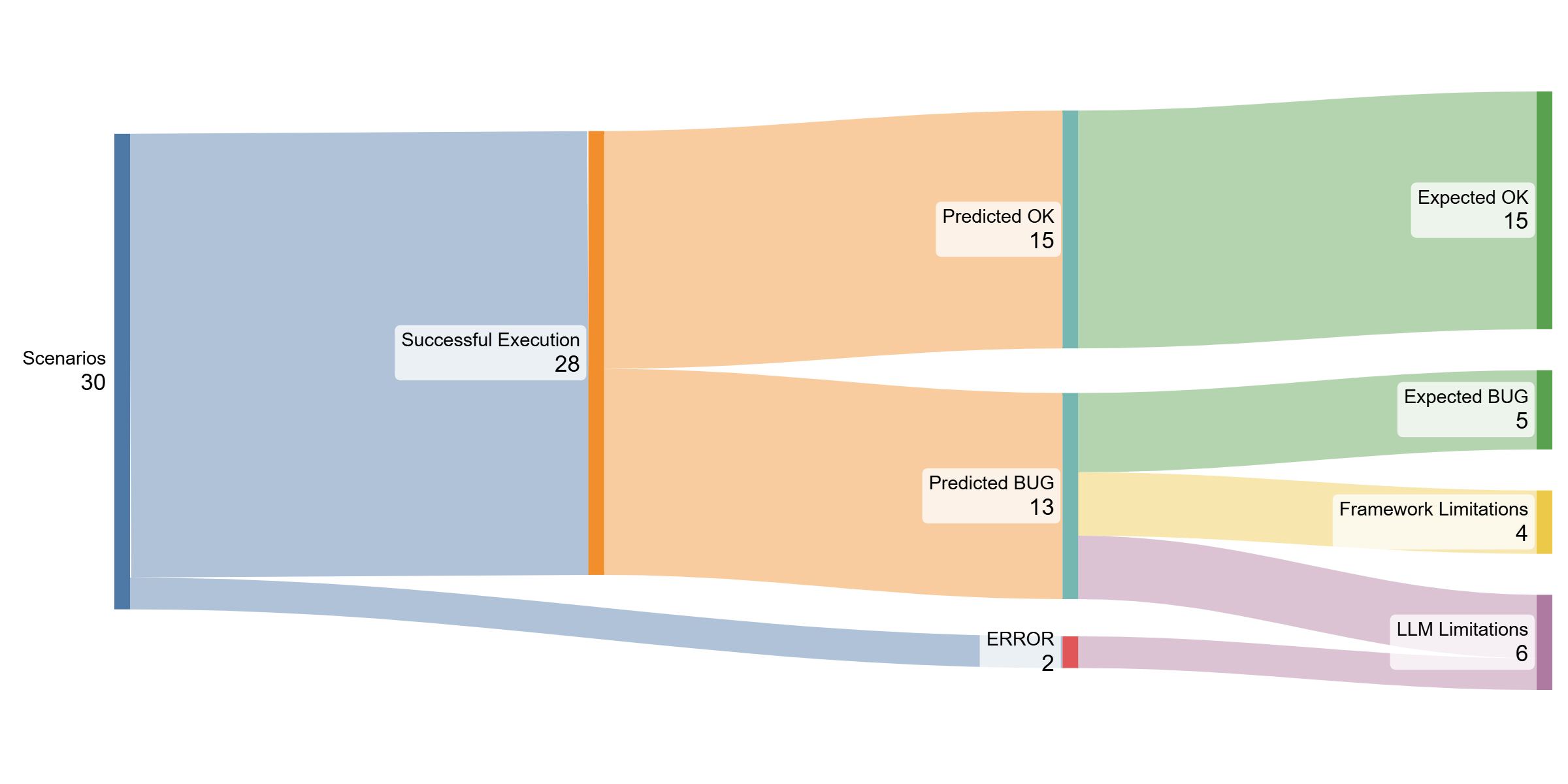}
    \caption{Outcome distribution across 30 evaluated scenarios}
    \label{fig:outcome_distribution}
\end{figure}

The outcome distribution across all thirty scenarios is depicted in Figure~\ref{fig:outcome_distribution}. All figures reported in this subsection are post-revisitation; the separation between first-pass and post-revisitation outcomes is analysed in Section~\ref{sec:ablation}. Under our revisitation protocol, a scenario that produced a BUG verdict on the first pass retains that verdict when the subsequent revisitation with the stronger model yields only an execution error, since revisitation is intended to resolve uncertain results rather than to discard verdicts already obtained. Of the 30 scenarios, 28 completed the full multi-agent pipeline and produced a classification (93.3\% completion), while the remaining 2 terminated prematurely due to execution errors; among the 28 completed, 15 were classified as OK and 13 as BUG.

Manual inspection against the ground truth revealed 15 scenarios correctly classified as OK and 5 correctly identified as genuinely faulty. The remaining 10 disagreed with the ground truth: 8 false positives, classified as BUG although their expected outcome was OK, and 2 error terminations. Their root causes split across two categories: 4 attributable to LLM limitations and 4 to framework interaction limitations, such as unsupported gestures and transient UI states.

\begin{table}[htbp]
\centering
\begin{tabular}{lcc}
& \textbf{Predicted OK} & \textbf{Predicted BUG} \\
\textbf{Expected OK} & 15 & 8 \\
\textbf{Expected BUG} & 0 & 5 \\
\end{tabular}
\caption{Confusion matrix for fault detection (28 completed scenarios). All five genuinely faulty scenarios were detected as defects.}
\label{tab:confusion-matrix}
\end{table}

\begin{table}[htbp]
\centering
\begin{tabular}{ll}
\textbf{Metric} & \textbf{Value} \\
Pipeline Completion Rate & 28/30 = 93.3\% \\
Fault Detection Recall & 5/5 = 100\% \\
Fault Detection Precision & 5/13 = 38.5\% \\
Accuracy (completed scenarios) & 20/28 = 71.4\% \\
\end{tabular}
\caption{Summary of evaluation metrics.}
\label{tab:metrics}
\end{table}

\subsection{Single-Agent Baseline}\label{sec:ablation}

To justify the multi-agent decomposition and provide a baseline for interpreting the results, the same 30 scenarios were executed using a single-agent configuration in which a single LLM agent is responsible for all cognitive subtasks: visual perception, action planning, element identification, coordinate resolution, and outcome evaluation. Both configurations used \texttt{claude-sonnet-4.6} as the underlying model and followed the same revisitation procedure with \texttt{claude-opus-4.6}. Table~\ref{tab:firstpass} compares both pipelines, contrasting their first-pass and post-revisitation outcomes.

In the single-agent pipeline, one agent session alternates two turns per interaction step. In the \textit{action turn}, the agent receives the current screenshot, the full filtered XML page source, and a resource identifier reference file, and returns a single JSON response selecting a UI element by its resource identifier, the action type (tap, swipe, or text input), and the intended interaction; an invalid identifier triggers a re-prompt with that identifier banned, analogous to the Translator's retry mechanism. In the \textit{evaluation turn}, it receives the resulting screenshot and classifies the outcome as Finished, Correct, or Incorrect. Both turns share the same conversation context, so the full message history accumulates throughout the scenario. The deterministic coordinate lookup and device execution are shared with the multi-agent pipeline, ensuring any performance difference is attributable solely to the agent decomposition rather than to the device interface.

\begin{figure}[htbp]
\centering
\begin{tikzpicture}
\begin{axis}[
    ybar,
    width=0.85\textwidth,
    height=5.2cm,
    bar width=16pt,
    ymin=0, ymax=82,
    ylabel={First-pass value (\%)},
    symbolic x coords={Accuracy, Precision, FP rate},
    xtick=data,
    enlarge x limits=0.35,
    legend style={at={(0.5,-0.16)},anchor=north,legend columns=-1},
    nodes near coords,
    nodes near coords style={font=\footnotesize,/pgf/number format/fixed,/pgf/number format/precision=1},
]
\addplot coordinates {(Accuracy,40.0) (Precision,21.7) (FP rate,72.0)};
\addplot coordinates {(Accuracy,54.5) (Precision,23.1) (FP rate,52.6)};
\legend{Single-Agent, Multi-Agent}
\end{axis}
\end{tikzpicture}
\caption{First-pass performance of both pipelines under the same inexpensive \texttt{claude-sonnet-4.6} model. Higher is better for accuracy and precision; for the false-positive rate lower is better. The multi-agent decomposition substantially reduces the first-pass false-positive rate while improving accuracy, extracting more reliable verdicts from a weaker, cheaper model.}
\label{fig:firstpass}
\end{figure}
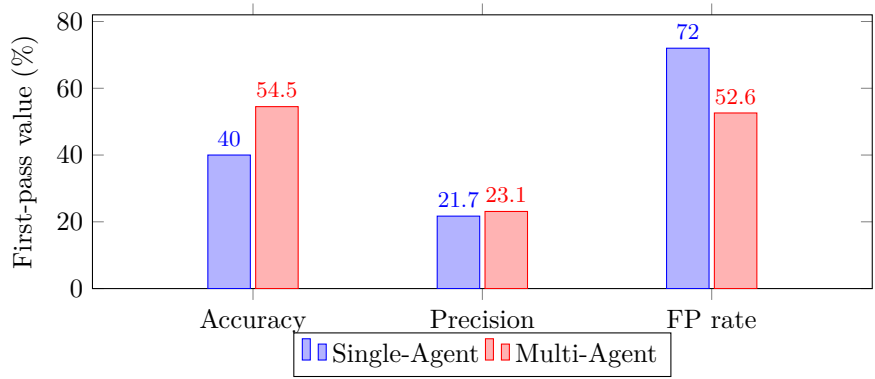

After revisitation, the two configurations converge to identical aggregate performance: both complete 28 of 30 scenarios, reaching 71.4\% completed-scenario accuracy, 38.5\% precision, and 5/5 fault-detection recall. This endpoint equivalence, however, masks a critical difference in first-pass behavior that the revisitation mechanism conceals (Table~\ref{tab:firstpass}, Figure~\ref{fig:firstpass}). On its initial execution with \texttt{claude-sonnet-4.6} alone, the single-agent classified 23 of 30 scenarios (76.7\%) as BUG, producing 18 false positives with a precision of only 21.7\% and an accuracy of 40.0\%. In contrast, the multi-agent pipeline's first pass classified only 13 of 30 as BUG (43.3\%) with 10 false positives, achieving 23.1\% precision and 54.5\% accuracy among the 22 scenarios it completed. The multi-agent pipeline thus commits far fewer false positives on the first pass, but it also terminates more scenarios with errors (8 versus 0): rather than forcing a potentially incorrect verdict, its Translator and Inspector retry mechanisms exhaust their budget and abort, converting some would-be false positives into explicit errors that are trivially triaged. Because both pipelines use the same inexpensive first-pass model, this reduction is obtained without any increase in model tier: the multi-agent decomposition extracts more reliable verdicts from a weaker, cheaper model, reducing reliance on costly stronger-model revisitation, on which the single-agent pipeline depends far more heavily.

\begin{table}[htbp]
\centering
\begin{tabular}{lcccc}
& \multicolumn{2}{c}{\textbf{Single-Agent}} & \multicolumn{2}{c}{\textbf{Multi-Agent}} \\
\textbf{Metric} & \textbf{First pass} & \textbf{Final} & \textbf{First pass} & \textbf{Final} \\
\hline
Pipeline Completion & 30/30 & 28/30 & 22/30 & 28/30 \\
Classified OK & 7 & 15 & 9 & 15 \\
Classified BUG & 23 & 13 & 13 & 13 \\
Terminated ERROR & 0 & 2 & 8 & 2 \\
True Positives & 5 & 5 & 3 & 5 \\
False Positives & 18 & 8 & 10 & 8 \\
Fault Recall & 5/5 & 5/5 & 3/5 & 5/5 \\
Accuracy (completed) & 40.0\% & 71.4\% & 54.5\% & 71.4\% \\
Precision & 21.7\% & 38.5\% & 23.1\% & 38.5\% \\
\end{tabular}
\caption{First-pass (\texttt{claude-sonnet-4.6}) versus post-revisitation (\texttt{claude-opus-4.6}) outcomes for both pipelines on the same 30 scenarios. Fault recall is reported over all five known defects; the two configurations converge to identical final performance but diverge sharply on the first pass.}
\label{tab:firstpass}
\end{table}

Revisitation was applied to every scenario that concluded the first pass with a BUG or ERROR status. In the multi-agent pipeline, 21 scenarios were revisited and seven changed classification, all of them improvements: six were resolved to OK and one genuine defect that had error-terminated on the first pass was correctly recovered as BUG, while the remaining revisited scenarios retained their first-pass verdict. Revisitation was not strictly monotonic, as with four scenarios the stronger model regressed a first-pass BUG verdict into an execution error, and under the retention protocol described in Section~\ref{sec:validation} these keep their first-pass BUG classification, which is why they are counted among the 13 final BUG outcomes rather than as errors. In the single-agent pipeline, all 23 first-pass BUG scenarios were revisited: 8 were resolved to OK, 2 regressed to execution errors, and 13 remained classified as BUG. The heavier reliance of the single-agent pipeline on revisitation, resolving 8 of its 23 flagged scenarios, quantifies how much of its final accuracy is contributed by the stronger model rather than by its own first-pass reasoning.

The single-agent's first-pass over-reporting stems from conflating navigational difficulty with system failure: lacking structural separation, when it cannot interact with an element as expected or reaches an unplanned screen state, it cannot distinguish its own misnavigation from a genuine fault and defaults to classifying a defect. A revealing failure mode was self-contradiction between the action and evaluation turns, since the agent would correctly execute an action, then, with the full conversation context accumulated, misattribute what it had just done and conclude that the system misbehaved, a context-management failure largely absent under the stronger revisitation model. The multi-agent pipeline structurally eliminates this, as the Evaluator sees only the before and after screenshots and the action description, without the accumulated navigational history, and the Translator's retry mechanism absorbs element-identification errors, the Inspector catches incorrect mappings before interaction, and the Evaluator assesses outcomes independently of the preceding navigation. This decomposition reduces the first-pass false positive rate, defined over the non-faulty scenarios that actually produced a verdict, from 72.0\% (18/25) for the single-agent pipeline to 52.6\% (10/19) for the multi-agent pipeline. The two denominators differ because each counts only the expected-OK scenarios that the respective pipeline classified: the single-agent pipeline produced no first-pass errors and therefore returned a verdict on all 25 non-faulty scenarios (18 false positives and 7 true negatives), whereas the multi-agent pipeline aborted six non-faulty scenarios into explicit errors and classified only the remaining 19 (10 false positives and 9 true negatives). Far from inflating the comparison, this makes it conservative, since the scenarios the multi-agent pipeline converted into errors are precisely those most likely to have become false positives had a verdict been forced, so its rate is computed over a harder residue. The reduction thus provides meaningful mitigation against the hallucination-prone behavior that reviewers of single-agent approaches have identified as a fundamental limitation.

\subsection{Run-to-Run Variance}\label{sec:variance}

To quantify the nondeterministic nature of LLM-based execution, a subset of six scenarios was selected for repeated execution, each run five times under identical conditions using \texttt{claude-sonnet-4.6}. The selection spans the full range of observed scenario complexity, from simple single-step verifications to multi-step deep menu navigations, and includes both correctly-functioning scenarios and a known system defect. Table~\ref{tab:variance} summarizes the per-scenario outcome distributions across the five runs.

\begin{table}[htbp]
\centering
\begin{tabular}{lcccccc}
\textbf{Scenario} & \textbf{GT} & \textbf{OK} & \textbf{BUG} & \textbf{ERR} & \textbf{Modal} & \textbf{Avg Calls} \\
\hline
Status bar & OK & 5 & 0 & 0 & OK & 17.6 $\pm$ 1.4 \\
Bluetooth toggle & OK & 5 & 0 & 0 & OK & 26.2 $\pm$ 5.9 \\
System settings & OK & 4 & 0 & 1 & OK & 40.6 $\pm$ 5.0 \\
Seats app search & OK & 4 & 0 & 1 & OK & 41.6 $\pm$ 9.2 \\
Wi-Fi Access Point toggle & OK & 1 & 2 & 2 & --- & 79.6 $\pm$ 5.5 \\
Language selection & BUG & 0 & 5 & 0 & BUG & 66.6 $\pm$ 18.0 \\
\end{tabular}
\caption{Outcome distribution across 5 repeated runs per scenario. GT = ground truth. Avg Calls = mean $\pm$ standard deviation of LLM invocations per run.}
\label{tab:variance}
\end{table}

Three of the six scenarios exhibited perfect stability, producing identical outcomes across all five runs: the status bar and Bluetooth toggle scenarios consistently classified as OK, and language selection consistently detected the known defect (BUG). Two further scenarios (system settings and the Seats app search) were mostly stable, producing the correct modal outcome in four of five runs, with a single spurious ERROR in one run each attributable to a translation retry limit being reached. The remaining scenario targeting the Wi-Fi Access Point toggle, which requires deep menu navigation through connectivity settings, was the most complex in terms of required interactions and exhibited the highest variance, with outcomes distributed across all three categories.

Across all 30 executions, 25 (83.3\%) agreed with the modal result for their scenario and 24 (80.0\%) matched the ground truth. Notably, fault detection was perfectly stable: the known defect was flagged in all five repetitions. The results reveal a clear correlation between scenario complexity and outcome variance, as scenarios requiring few navigation steps (one to three) were deterministic while those requiring extensive multi-step navigation (over seven) exhibited greater instability, consistent with the accumulation of stochastic decision points. Importantly, the dominant source of variance is not classification disagreement but pipeline completion: when a scenario reaches a verdict (OK or BUG), that verdict is consistent across runs with the same model, and the variance instead manifests as some runs terminating with an ERROR due to translation retry exhaustion or step limit violations, without ever producing a conflicting behavioral classification.

\subsection{Computational Cost and Usage}\label{sec:usage}

Because operational cost and latency are decisive for pipeline integration, every LLM invocation was instrumented and token consumption, call counts, and monetary cost were aggregated for both pipelines. Cost is expressed in premium requests, the billing unit of the LLM service used, where each call is weighted by model (one for \texttt{claude-sonnet-4.6}, three for \texttt{claude-opus-4.6}) and converted to an estimated monetary figure at \$0.04 per premium request. Cached prompt re-sends, which dominate raw input-token counts because the multimodal system prompts and prior screenshots are resubmitted at every step, are billed separately at a negligible rate and excluded from the fresh-input figures (Table~\ref{tab:usage}).

\begin{table}[htbp]
\centering
\begin{tabular}{lcccccc}
& \multicolumn{3}{c}{\textbf{Multi-Agent}} & \multicolumn{3}{c}{\textbf{Single-Agent}} \\
\textbf{Metric} & \textbf{First} & \textbf{Revisit} & \textbf{Total} & \textbf{First} & \textbf{Revisit} & \textbf{Total} \\
\hline
LLM calls & 1{,}218 & 851 & 2{,}069 & 638 & 622 & 1{,}260 \\
Premium requests & 1{,}218 & 2{,}263 & 3{,}481 & 638 & 1{,}616 & 2{,}254 \\
Est. cost (USD) & 48.72 & 90.52 & 139.24 & 25.52 & 64.64 & 90.16 \\
Output tokens & 162.5k & 136.8k & 299.3k & 161.1k & 158.6k & 319.7k \\
Fresh input tokens & 2.89M & 2.20M & 5.09M & 2.16M & 3.33M & 5.49M \\
Cost per scenario (USD) & --- & --- & 4.64 & --- & --- & 3.01 \\
\end{tabular}
\caption{Computational cost and usage for both pipelines over 30 scenarios, split into first-pass (\texttt{claude-sonnet-4.6}) and revisitation (\texttt{claude-opus-4.6}). Premium requests weight each call by model (Sonnet $\times1$, Opus $\times3$); estimated cost assumes \$0.04 per premium request. Fresh input excludes cached prompt re-sends.}
\label{tab:usage}
\end{table}

The complete multi-agent run consumed 2{,}069 LLM calls and an estimated \$139.24 across both passes, of which the first pass with \texttt{claude-sonnet-4.6} accounted for 1{,}218 calls (\$48.72) and the revisitation pass with \texttt{claude-opus-4.6} for the remaining 851 calls (\$90.52). The revisitation pass, although covering only 21 scenarios, therefore dominates cost, both because \texttt{claude-opus-4.6} is billed at three times the per-call rate and because the hardest scenarios require the most steps. Averaged over all 30 scenarios, the whole pipeline costs approximately \$4.64 per scenario; the first pass alone averages 40.6 LLM calls per scenario, ranging from 17 for a single-step status-bar check to 98 for a deep multi-step navigation, at roughly \$1.62 per scenario. Wall-clock execution time, measured from scenario logs, averaged 8.4 minutes per scenario on the first pass with a median of 5.2, ranging from 1.5 to 48.7 minutes, confirming that per-scenario latency is dominated by the sequential agent round trips and the fixed three-second settling delay after each interaction, and that ARIA is suited to nightly regression runs rather than in-loop feedback.

Within a first-pass scenario, the four in-loop agents plus the summarizer are not invoked uniformly. The Translator is by far the most frequently called agent, accounting for 480 of the 1{,}218 first-pass calls (39.4\%), because each element-identification retry and each Inspector rejection triggers an additional Translator invocation. The Executor (195 calls, 16.0\%), Summarizer (193, 15.8\%), Evaluator (182, 14.9\%), and Inspector (168, 13.8\%) are invoked in comparable proportions. This distribution indicates that element resolution, rather than high-level planning or evaluation, is the principal cost driver and the most promising target for optimization.

The single-agent baseline is markedly cheaper, consuming 1{,}260 calls and an estimated \$90.16 in total (\$25.52 first pass, \$64.64 revisitation), roughly \$3.01 per scenario, because it invokes a single model per turn rather than four specialized agents per step. This confirms that the multi-agent decomposition trades a higher token and monetary budget for its reduction in first-pass false positives, a trade-off that the short-circuit replay mode is designed to amortize over successive software versions by replacing the full four-agent loop with a single Evaluator invocation per scenario during regression re-runs.

\subsection{Answers to Research Questions}

\textbf{RQ1: How can LLM-powered multi-agent architectures be designed to autonomously execute end-to-end test scenarios on automotive infotainment systems through visual interface interaction?}
The 93.3\% pipeline completion rate demonstrates that the four-agent decomposition (Executor, Translator, Inspector, Evaluator) is operationally viable for autonomous test execution on a real infotainment system: in 28 of 30 scenarios the closed loop iterated without human intervention across multi-step navigation, text input, toggle interactions, and scroll gestures. The separation between the Executor (visual reasoning only) and the Translator (XML-based element resolution) provided a clean boundary between perceptual reasoning and implementation-level interaction. The single-agent baseline (Section~\ref{sec:ablation}) gives empirical support: without the decomposition, a single agent using the same model classified 76.7\% of scenarios as defective on its first pass, versus 43.3\% for the multi-agent pipeline, showing that concentrating all cognitive responsibilities in one reasoning chain leads to systematic over-reporting of false positives. Agent independence kept the Executor's user-centric perspective uncontaminated by system-level knowledge, while the enforced JSON format and automatic re-prompting proved essential for reliable inter-agent communication.

\textbf{RQ2. To what extent can a vision-based, natural-language-driven testing framework reliably detect genuine system faults while minimizing false positives in an industrial infotainment validation environment?}

ARIA detected all five scenarios with genuinely incorrect system behavior, classifying each as BUG and producing zero false negatives (recall 5/5): no genuine fault was ever passed as working. Three of the detected faults corresponded to the deliberately disabled climate control subsystem, whose absent functionality the Evaluator correctly observed in each case; one of these was counted as detected per the run-to-run reasoning in Section~\ref{sec:validation} after intermittently error-terminating. Another involved a language selection operation where no alternative language could be selected, which the framework correctly flagged as a missing state change. The last was a system-wide black screen, a transient anomaly unrelated to the functionality under test, demonstrating the framework's ability to detect system-level issues that a narrowly scoped scripted test would likely miss.

Among the 28 completed scenarios, 20 (71.4\%) were classified in agreement with the ground truth, with precision reaching 38.5\%. The 8 false positives among completed scenarios are predominantly attributable to framework interaction and LLM limitations rather than fundamental failures of the multi-agent reasoning approach. The low precision highlights the need to decrease the false positive rate, mainly through collecting the relevant context of the system's expected behavior and supplying it to the Evaluator agent so that it can more accurately assess the system's state after every interaction.

\textbf{RQ3. What are the primary failure modes, limitations, and architectural trade-offs involved in delegating GUI test execution to LLM agents operating through coordinate-based interaction primitives?}

Root cause analysis of all 8 non-correct scenarios revealed two distinct failure categories in equal proportion, as shown in Figure \ref{fig:outcome_distribution}. LLM limitations (50.0\%) manifested as the Executor navigating to incorrect menu paths or exhausting the step limit while searching for a target, the Translator repeatedly generating resource identifiers that the Inspector rejected, or the Executor misreading the effect of a status-bar interaction. These failures illustrate a fundamental trade-off of LLM-based execution: the flexibility that enables agents to reason about unfamiliar interfaces also permits them to generate plausible but incorrect action sequences.

Framework interaction limitations accounted for the other 50.0\% of non-correct outcomes, arising from unsupported gesture types such as long-press-and-drag, transient UI states that cannot be captured through discrete screenshots, and insufficient contextual knowledge about specific element interaction requirements.

A qualitatively distinct observation emerged from the two home-screen swipe-up scenarios. In the final, post-revisitation results both are correctly classified as OK, but on the first pass with the base model one was flagged as a defect and the other error-terminated. Manual investigation revealed that the intended navigation requires a specific motion trajectory that is not visually communicated to the user; the base model, reasoning as a simulated human user, struggled with exactly this gesture before the stronger revisitation model completed it. This first-pass friction, although resolved in the final tally and therefore not counted among the non-correct outcomes, indicates that real users may encounter the same difficulty and illustrates that vision-based agents can surface unintuitive interaction patterns that would be difficult to detect through traditional scripted automation.

\textbf{RQ4. What are the practical challenges and prerequisites for integrating LLM-based test execution frameworks into existing automotive validation workflows and continuous integration pipelines?}

The deployment of ARIA on the industrial test setup revealed several practical prerequisites for pipeline integration. The framework requires stable Appium connectivity with ADB-based device discovery; the checkpoint and manifest system mitigates network interruptions by enabling resumption from the last completed scenario, essential for unattended execution. Existing manual test scripts can be ingested through the scenario transformation mode without a device connection, though the quality of the generated scenarios directly impacts execution reliability and benefits from review.

The primary scheduling constraint is execution latency: four sequential agent invocations per interaction step, each transmitting multimodal data to external LLM services, yield far longer per-scenario times than scripted automation, positioning ARIA for nightly regression runs rather than rapid in-loop feedback. Token consumption from per-step screenshots translates directly into operational cost, partially mitigated by the short-circuit replay mode's single Evaluator invocation, while the model abstraction layer lets different LLM providers be substituted without modifying execution logic as availability or pricing changes.

\subsection{Threats to Validity}

\textbf{Internal validity.} The ground truth was established through manual inspection by a single reviewer with knowledge of the test setup's capabilities, introducing subjectivity in ambiguous cases such as the interpretation of the first-pass friction on the two swipe-up scenarios as a usability concern. The non-deterministic nature of LLM inference means that individual outcomes may not be perfectly reproducible across runs; however, the repeated execution experiment (Section~\ref{sec:variance}) demonstrates that 83.3\% of individual runs agree with the modal result, with perfect consistency for fault detection and simpler scenarios.

\textbf{External validity.} All experiments were conducted on a single physical infotainment test environment disconnected from a vehicle. The 30 scenarios evaluated  represent a fraction of the theoretically possible interaction sequences. With only 5 genuinely faulty scenarios, all of which were ultimately detected, the perfect recall figure (5/5) is derived from a small sample and should be interpreted with appropriate caution. The effectiveness of the framework at detecting regressions across software updates was not evaluated due to time constraints.

\textbf{Construct validity.} The Translator agent was provided with supplementary reference files mapping resource identifiers to UI element descriptions. While these do not expose internal APIs, they provide prior knowledge about the system's naming conventions, and their absence could reduce performance on undocumented systems.

\section{Related Work}\label{sec:rel_work}

The emergence of Large Language Models has opened new avenues for automating the execution of GUI-based test scenarios \cite{faraji_ai-driven_2025, sezgin_leveraging_2025}. Rather than encoding interaction logic programmatically, these approaches exploit the reasoning and perception capabilities of LLMs to navigate user interfaces and perform high-level tasks. Agent-based architectures are particularly prevalent, because GUI exploration without a separation of responsibilities is prone to failures, such as selecting actions unavailable in the current interface state, repeating ineffective interactions, or otherwise deviating from the intended exploration policy \cite{ran_guardian_2024}. By decomposing responsibilities such as planning, action execution, state observation, and reflection, these architectures externalize the control logic and state management that would otherwise be left to a single model, improving instruction adherence and reducing unproductive exploration loops.

\textbf{GUARDIAN} investigates LLM agents in automated feature-based UI testing for mobile applications \cite{ran_guardian_2024}. It addresses two challenges inherent to LLM-driven execution: the tendency to fail at following complex domain-specific instructions, and the inability to re-plan when exploration falls into repetitive loops. To mitigate these, GUARDIAN employs computation offloading, shifting state management and logical constraints from the LLM to an external orchestration layer, executing and validating a single UI action per round until the objective is reached.

\textbf{LLMDroid} takes a hybrid approach, enhancing traditional automated GUI testing tools by integrating LLM reasoning only when code-coverage growth slows \cite{wang_llmdroid_2025}. Rather than querying the model for every action, it relies on a traditional tool for high-speed ``Autonomous Exploration'' and transitions to an ``LLM Guidance'' stage that summarizes explored pages and selects target functionalities with the highest potential for discovering unexplored paths when coverage plateaus.

\textbf{GERALLT} uses two specialized LLM agents, a ``controller'' and an ``evaluator'', for automated GUI testing \cite{rosenbach_automated_2025}. The controller performs interactions to achieve loosely defined objectives from the current GUI state, documentation, and action history, while the evaluator analyzes before/after screenshots to detect unintuitive behaviors, visual inconsistencies, or functional errors. A GUI Parser translates visual elements into structured metadata, letting the agents reason about the interface without human interaction data.

The reviewed literature reveals several gaps that ARIA addresses. Firstly, most LLM-driven testing frameworks target mobile or web applications \cite{yoon_intent-driven_2024, zimmermann_gui-based_2023, hu_kuitest_2025, kapoor_ai-assisted_2025}; solutions applied to the automotive infotainment domain remain scarce and insufficient for the specific constraints of this environment \cite{wang_automating_2025, wynn-williams_can_2025, khaliq_deep_2022, s_m_n_s_k_seneviratne_ad-pu_2023}. 

Secondly, existing approaches such as GUARDIAN and LLMDroid predominantly employ single-agent or dual-agent architectures that concentrate multiple cognitive responsibilities (perception, planning, action selection, and validation) within one or two models \cite{ran_guardian_2024, wang_llmdroid_2025}. ARIA's four-agent decomposition pushes the multi-agent paradigm further by assigning each sub-task to a dedicated specialist, reducing brittleness and improving controllability compared to more monolithic approaches. Separating these responsibilities ensures, for instance, that an agent performing exploratory UI traversal cannot be influenced by information available to a separate agent responsible for classifying system behavior against specifications.

Thirdly, ARIA accepts existing manual test scripts as natural-language input, eliminating the need for test re-authoring or formalization into structured specifications, directly targeting the identified bottleneck: organizations do not lack test cases, they lack the capacity to execute them at the required pace.

Finally, ARIA's closed-loop interaction model enables the system to respond to the evolving interface state of the system under test. The translated interaction scripts are executed on the infotainment system; if no errors occur, a new state description is sent to the Executor agent for evaluation and subsequent action determination, while execution errors are relayed back to the Translator agent for correction before resubmission. By embedding evaluation from an end-user perspective after each action, the approach moves fault detection closer to the point of interaction, producing structured result reports, reproducible action scripts, and visual evidence for every executed step.

\section{Conclusion}\label{sec:conclusion}

It can be concluded that, by decomposing the test execution process into four specialized agents, each responsible for a distinct cognitive subtask, ARIA demonstrates that LLM-powered architectures can autonomously navigate, interact with, and evaluate complex infotainment UIs without human intervention and without coupling to internal system APIs or implementation details.

The industrial evaluation, conducted on a physical Android-based infotainment test environment, yielded encouraging but qualified results. ARIA completed the full execution pipeline for 93.3\% of the evaluated scenarios and, among completed scenarios, never passed a genuine fault as working, detecting all five known defects. The overall classification accuracy among completed scenarios reached 71.4\%, and precision remained low at 38.5\%, with the majority of non-correct outcomes attributable to agent-native and mechanical limitations of the interaction primitives rather than fundamental shortcomings of the multi-agent reasoning approach. A notable observation was that the home-screen swipe-up scenarios, although correctly classified once revisited, produced first-pass friction that mirrors a latent usability concern, demonstrating that vision-based agents simulating human user behavior can reveal unintuitive interaction patterns alongside software defects.

Several directions for future work emerge from these findings. The LLM limitations could be mitigated through richer contextual grounding, such as navigation maps or terminology glossaries, and self-verification mechanisms that compare the observed screen state against the agent's assumptions before proceeding. The framework limitations point to the need for extended interaction primitives supporting compound gestures such as long-press-and-drag and gestures anchored to element boundaries. The short-circuit replay mode, implemented but not evaluated for cross-version regression detection here, is a natural next step in a continuous integration setting with successive builds. Finally, the model abstraction layer positions ARIA to benefit from advances in LLM capabilities, as improvements in visual reasoning and instruction adherence should reduce the misclassification rate without architectural changes.

The results presented in this work demonstrate that LLM-powered multi-agent test execution is feasible in an industrial automotive context, while making clear that its current precision and error-rate leave meaningful work before unattended deployment in a continuous integration pipeline. By accepting natural-language scenario descriptions as input and producing structured reports, reproducible action scripts, and visual evidence for every step, ARIA is best positioned today as an early-stage, human-supervised aid for nightly regression campaigns, narrowing the gap between manual testing practices and the automated execution capacity that modern release cycles demand.

\section*{Data Availability}
The data supporting this paper were obtained under a non-disclosure agreement with a software company. This agreement prohibits the sharing of source code, internal and derived artifacts, and specific details about the testing equipment and the intricacies of the testing environment. We have provided sufficient architectural detail about the framework and the infotainment system requirements in Section \ref{sec:aria} to allow for replication in other organizational contexts.



\bibliography{lipics-v2021-sample-article}

\appendix

\end{document}